\documentclass[superscriptaddress,amsmath,amssymb,aps,pra,nofootinbib]{revtex4}
\usepackage{graphicx}

\usepackage{tikz}

\begin{document}

\title{Discovering nonlinear resonances through physics-informed machine learning}

\author{G. D. Barmparis}

\affiliation{Institute of Theoretical and Computational Physics and Department of Physics, University of Crete,  P.O. Box 2208, 71003 Heraklion, Greece\\ }
\affiliation{National University of Science and Technology MISiS,  Leninsky prosp. 4, Moscow 119049, Russia}

\author{G. P. Tsironis}

\affiliation{Institute of Theoretical and Computational Physics and Department of Physics, University of Crete,  P.O. Box 2208, 71003 Heraklion, Greece\\ }
\affiliation{National University of Science and Technology MISiS,  Leninsky prosp. 4, Moscow 119049, Russia}

\date{\today}

\begin{abstract}
For an ensemble of nonlinear systems that model for instance molecules or photonic systems we propose a method that finds efficiently the configuration that has prescribed transfer properties.  Specifically,  we use physics-informed machine-learning (PIML) techniques to find the parameters for the efficient transfer of an electron (or photon) to a targeted state in a non-linear dimer. We create a machine learning model containing two variables, $\chi_D$ and $\chi_A$, representing the non-linear terms in the donor and acceptor target system states. We then introduce a data-free physics-informed loss function as $1.0 - P_j$, where $P_j$ is the probability, the electron being in the targeted state, $j$. By minimizing the loss function, we maximize the occupation probability to the targeted state. The method recovers known results in the Targeted Energy Transfer (TET) model and it is then applied to a more complex system with an additional intermediate state.  In this trimer configuration the PIML approach discovers desired resonant paths from the donor to acceptor units. The proposed PIML method is general and may be used in the chemical design of molecular complexes or engineering design of  quantum or photonic systems.

\end{abstract}

\maketitle

\section{Introduction}

Consider an ensemble of physical systems that share common dynamics, classical or quantum. The systems are similar, in the sense that they differ only in the actual values of one or more parameters. For instance, assume one dimensional motion of a single particle in a quartic potential $V(x) = \alpha x^2 + \beta x^4$, where $x$ is the particle position.  Each set of values of the parameters $( \alpha ,~\beta )$ label a different physical system and while all members of this ensemble share the same equations of motion the specific dynamics is evidently different.  While small changes to the parameters may, in general,  not have drastic changes in the evolution, in other cases the dynamics can be extremely different.  If, for instance we take positive values for the two parameters we have bounded motion, while for the same absolute values but negative we have completely unbounded motion.  In nonlinear systems, the specific dynamics depend, additionally, to the initial conditions.  For simplicity of the arguments that follow, lets us fix the initial conditions to be the same for all dynamical systems of the ensemble.  Assume now we have a subset of systems that follow an additional rule imposed externally.  This rule can be simple or more complicated; for instance consider the subclass of systems that traverse a certain point with a given momentum or a range of values in momentum.  If there are many members in the subset we may seek a specific set defined again according to certain criteria.  Let us call the values of the parameters that follow the externally imposed rule $( \alpha_N ~, \beta_N )$.

We view the process of finding this desired dynamical system as an evolutionary procedure from $(\alpha_0 , \beta_0 )$, to $(\alpha_1 , \beta_1 )$ to ... $( \alpha_N ~, \beta_N )$.  We start from a (random) set of initial parameters, follow the dynamics and evaluate the criterion we seek to impose. Since the criterion almost surely is not satisfied we back-propagate the error and find a set of new values of the parameters that will bring us closer to the desired constraint.  We iterate until we obtain the system that is as close as possible to the desired constraint. This procedure gives through reversed engineering the system that fulfills the criteria set in an almost algorithmic fashion.  In some sense, it allows us to discover properties within the ensemble of systems that, otherwise would need cumbersome and tedious analytical and numerical tools.

Let us fix ideas with a specific, simple example.  We consider  the discrete Nonlinear Schr{\"o}dinger equation (DNLS)  that is ubiquitus in nonlinear physics with numerous applications from condensed matter physics, to optics,  photonics and biological physics.  We may write it in a Hamiltonian form

\begin{equation}\label{EQ-1}
H = \sum_n \left[ \frac{\epsilon_n}{2} |\psi_n |^2 +  \frac{\chi_n }{4} |\psi_n |^4  + V \left(
\psi_{n}\psi_{n+1}^* + \psi_{n+1}^* \psi_n \right)\right]
\end{equation}

where $\psi_n \equiv \psi_n (t)$ is a complex variable at time $t$ while $\epsilon_n $, $\chi_n$ and $V$ are parameters of the problem.  It is more intuitive to cast this problem into a nonlinear tight-binding problem though the use of Hamilton equations on Eq. (\ref{EQ-1}) \cite{eilbeck}; we obtain

\begin{equation}\label{EQ-2}
i \frac{d \psi_n}{dt} = \epsilon_n \psi_n + V( \psi_{n+1} + \psi_{n-1} ) + \chi_n |\psi_n |^2 \psi_n
\end{equation}

In this representation we can think of a one-dimensional infinite lattice where each site is labeled by the index $n$, we have local site energy $\epsilon_n$ and local non-linearity $\chi_n$ while $V$ is the common nearest-neighbor integral overlap.  A quantum mechanical particle tunnels from site to site while experiencing a nonlinear interaction due to strong coupling with other degrees of freedom-in the LHS of the equation we have suppressed $\hbar$.  This DNLS equation has been applied to numerous problems in physics and mathematics.  We focus at first in the simplest possible configuration of the nonlinear dimer,  i.e. a system with only two sites, a Donor - Acceptor dimer.  For the sake of simplicity we choose the initial condition where the particle is initially on the first site (Donor) and search for the time evolution of the occupation probabilities.  We may define the energy difference $\epsilon = \epsilon_D - \epsilon_A$ and thus have the following parameter set $\vec{p}= ( \epsilon , V, \chi_D , \chi_A )$. Although the transfer parameter $V$ may be eliminated and turn the remaining three parameters in reduced ones, we prefer to keep it for intuitive reasons.  The degenerate dimer problem with equal non-linearities, i.e. $\chi_D = \chi_A$, is solvable in terms of Jacobian elliptic functions while the non-degenerate one in terms of Weierstrass elliptic functions \cite{kenkre, gpt1, christodoulides, gpt2}.  It is known analytically that the fastest transfer from one site to the second one is in the degenerate linear case, i.e. for $\epsilon = 0$ with $\chi \equiv \chi_D = \chi_A =0$; in this case the transfer period $T$ for the localized initial condition we consider is $T_0 =\pi / 2V$.  In the presence of non-linearity ($ \chi < 4 V$) this period grows as  $T=T_0 K( \chi / 4 V )$, where $K$ is the complete elliptic integral of first kind,  while  for larger non-linearity values and/or non-degeneracies the transfer is incomplete.  The reduction of the transfer due to non-linearity may be considered as a signature of polaronic effects.

We may now pose the following question: Within the class of nonlinear dimers labeled by the set $\vec{p}$, which is the one that has the desired transfer from site one to site two in addition to the linear, degenerate one ($\epsilon = \chi_D = \chi_A = 0$)?  One obvious way to proceed is first to find the general solution of the dimer problem for arbitrary $\vec{p}$, calculate the oscillation period and find the set of the desired parameters.  Clearly this is a very cumbersome if not impossible procedure.  An alternative way is to use the form of back-propagation described previously.  The specific model of desired transfer, or targeted energy transfer (TET), was introduced and analyzed several years ago with nonlinear dynamics techniques \cite{kopidakis1,kopidakis2}.  The aim of the present work is to cast the model into a ML framework and find its solution with alternative techniques.  Once this is accomplished, the methodology will be applied to a non-analytically solvable problem and thus obtain new results.  The structure of this paper is then the following.  In the next section II, we introduce the nonlinear dimer model through its dynamical equations and describe the PIML motivated methodology we use.  The application of the latter leads to the designed transfer result of TET.  In section III, we apply this methodology to a trimer model that consisted of the Donor-Acceptor dimer unit separated by an additional intermediate linear unit, $\Phi$, that plays the role of a barrier. Application of the PIML method in this trimer leads to discovering a specific set of parameters resulting in the most efficient transfer configuration.  Finally in section IV, we discuss our findings and describe the outlook of the method.

\section{Nonlinear dimer}
\subsection{Dynamical equations}

The DNLS equation describes equally well photonic and molecular systems.  In photonics it describes the propagation in space of a mode in a fiber while the latter may interact with other fibers that may or not include nonlinear susceptibilities \cite{hennig}. In chemical physics it describes molecular units where an electron or excitation propagates in time; the latter  may transfer to an alternative unit while the nonlinear term incorporates the interaction with phonons in an approximate way.  To simplify our presentation we will use the ``chemical physics" language for DNLS but we will give also conclusions in the the ``photonics" interpretation.

\begin{figure}[ht]
\includegraphics[width=6cm]{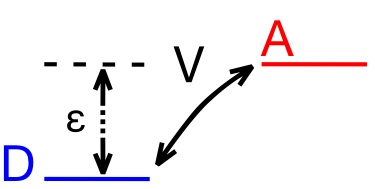}
\caption{Nonlinear dimer. Energy levels.}
\label{nl-dimer-sketch}
\end{figure}

In chemical physics language thus we consider a dimer system with two energy levels as shown in Fig (\ref{nl-dimer-sketch}). The energy difference between the two levels,  $\epsilon$, and  the transition matrix element, $V$, define the electronic or exitonic part of the system, while additionally, the nonlinear term included represents polaronic or other self-interacting effects. Using the nomenclature of TET we call the first unit {\it donor} while the second {\it acceptor} \cite{kopidakis2}.  

The donor-acceptor (D-A) system is described through the nonlinear Scr\"odinger equation that is specialized as follows:
\begin{eqnarray}
\begin{split}
i \dot{\psi}_{D} & = & -\epsilon \psi_D + V \psi_A + \chi_D | \psi_D |^2 \psi_D \\
i \dot{\psi}_{A} & = & V \psi_D + \chi_A | \psi_A |^2 \psi_A
\end{split}
\label{nl-schrodinger}
\end{eqnarray}
where, $\psi_D$, and $\psi_A$, are the wave-functions in first (donor) and the second (acceptor) dimer sites, respectively. This D-A unit has, in addition to $\epsilon$ and $V$ two nonlinearity parameters $\chi_D$ and $\chi_A$ at the donor and acceptor sites respectively that, in general, are different.  The problem posed in TET and analyzed here is to find the set of parameters $\vec{p}$ that lead to the desired transfer of energy from D to A units.  The latter is measured by $ |\psi_A (t) |^2 $, i.e.  the occupation probability in the acceptor site given that initially the the particle was fully in the donor, i.e. $|\psi_D (0)|=1$. This problem of efficient transfer is related to ultra-fast processes in bio-molecular systems, such as the exciton transfer in chlorophyl \cite{aubry1}. In the following subsection we introduce the new methodology for the solution of this problem.
\begin{figure}[ht]
\includegraphics[width=9cm]{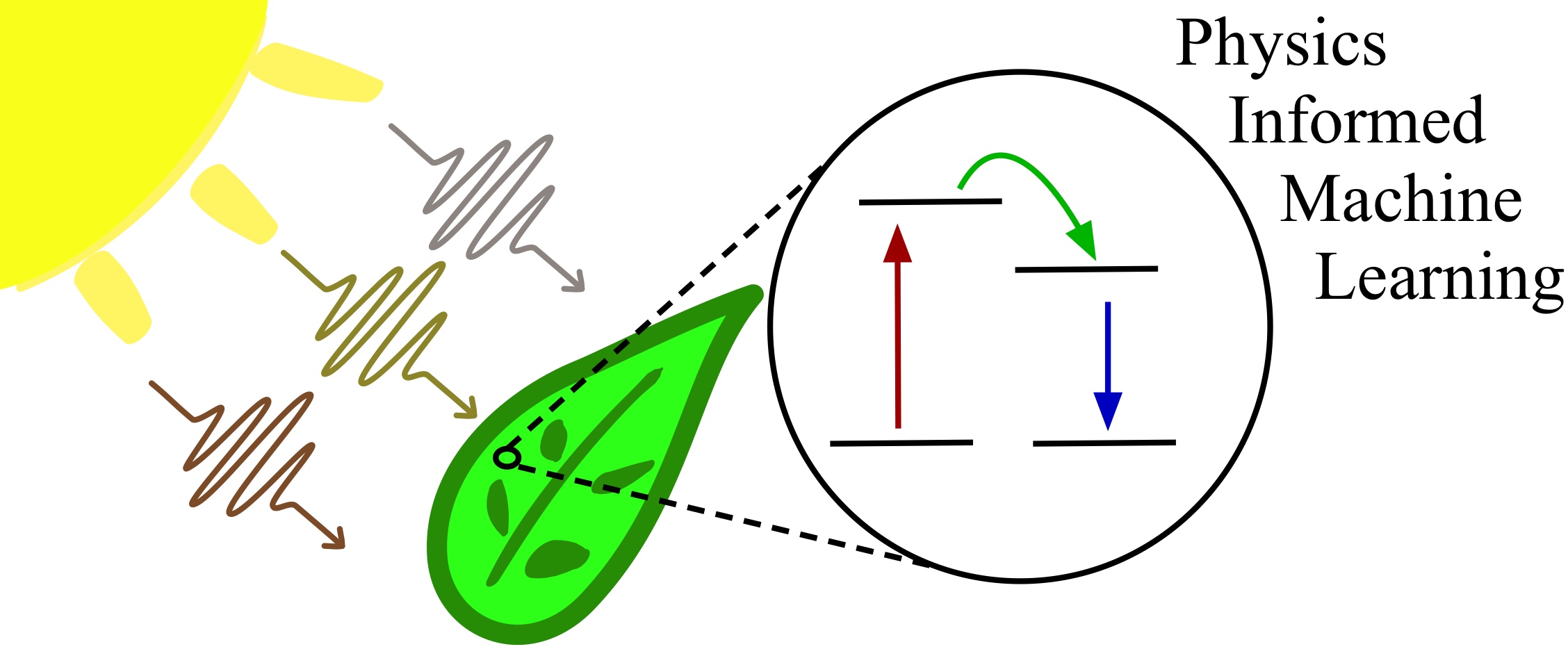}
\caption{Physics-Informed Machine Learning in the quantum realm can efficiently find configurations that conclude to designed exciton transfer in biomolecular systems.}
\label{tet}
\end{figure}
%

\subsection{Methodology}
The methodology we introduce is motivated by Physics-Informed Machine Learning (PIML) \cite{karniadakisNature, gdbSIR,pouliasi}, an approach that uses AI-based data techniques together with the dynamical equations of the phenomenon it aims to investigate. In a PIML framework, one can either create a model that replaces the physical system, for example, a neural network that, in addition to a data-driven model, learns to solve a given partial differential equation  \cite{karniadakis} or create a model that uses physics to learn the parameters that govern a system \cite{gdbSIR}. The latter methodology solves an inverse problem to find the values of the parameters that conclude to desired physical properties. In this case, the trainable parameters obtain physical meaning since they represent the physical quantity that governs the properties of the system. Thus, in a PIML algorithm, the imposed physical properties guide the process of discovering the values of the parameters that give to the system the desired properties. In our work, the required physical property of the system (resonant transfer) is governed by the value of the non-linear terms, and thus they take the role of the trainable parameters. 

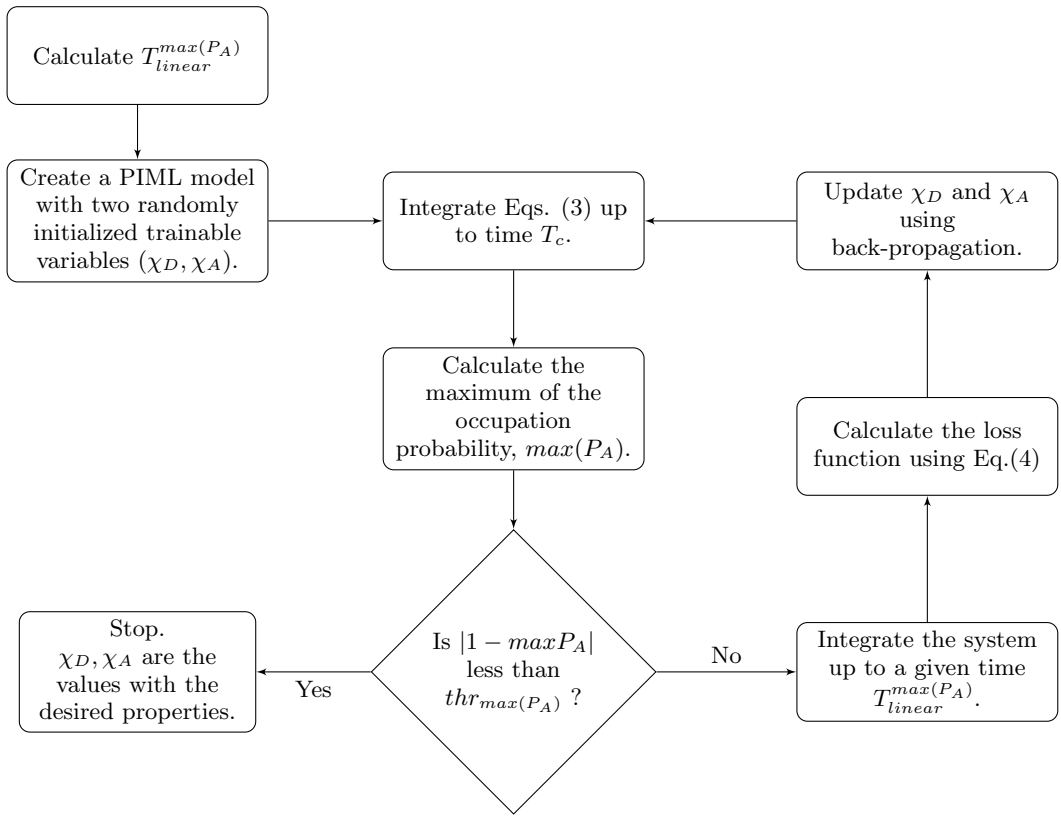
\begin{figure}[ht]
\usetikzlibrary{shapes.geometric, arrows, calc}

\tikzstyle{decision} = [diamond, draw, text width=4.5em, text badly centered, node distance=3cm]
\tikzstyle{block} = [rectangle, draw, text width=10em, text badly centered, rounded corners, minimum height=4em]  
\tikzstyle{line} = [draw, -latex']
\tikzstyle{terminator} = [ draw, ellipse, node distance=3cm, minimum height=2em]  

\begin{tikzpicture}[node distance=2cm, auto]  
  \node [block] (tLinear)  {Calculate $T^{max(P_A)}_{linear}$};
  \node [block, below of = tLinear, yshift = -2mm] (model)  {Create a PIML model with two randomly initialized trainable variables ($\chi_D, \chi_A$).};
  \path [line] (tLinear) -- (model);

  \node [block, right of = model, xshift = 30mm] (integrate)  {Integrate Eqs. (\ref{nl-schrodinger}) up to time $T_c$.} ; 
  \path [line] (model) -- (integrate);  

  \node [block, below of = integrate, yshift = -5mm] (probability)  {Calculate the maximum of the occupation probability, $max(P_A)$.};
  \path [line] (integrate) -- (probability);

  \node [decision, below of = probability, text width=7em, yshift = -5mm] (check) {Is $|1 - max{P_A}|$ less than $thr_{max(P_A)}$ ?};
  \path [line] (probability) -- (check);
  
  \node [block, text width=9em, left of = check, xshift = -30mm] (stop) {Stop. \\ $\chi_D, \chi_A$ are the values with the desired properties.};
  \path [line] (check.west) -- node {Yes} (stop.east);

  \node [block, right of = check, xshift = 35mm] (giventime)  {Integrate the system up to a given time $T^{max(P_A)}_{linear}$.};
  \path [line] (check.east) -- node[anchor=south] {No} (giventime.west);

  \node [block, above of = giventime, yshift = 10mm] (loss)  {Calculate the loss function using Eq.(\ref{loss})};
  \path [line] (giventime) -- (loss);
  
  \node [block, above of = loss, yshift = 10mm] (update)  {Update $\chi_D$ and $\chi_A$ using back-propagation.};
  \path [line] (loss) -- (update);
  
  \path [line] (update) -- (integrate);
        
\end{tikzpicture}
\caption{Flowchart portraying the AI algorithm used in the search for the nonlinear resonant configuration with the desired properties.}
\label{flowchart}
\end{figure}

We first solve numerically, by means of the $4^{th}$ order Runge-Kutta method (integration step = 0.001),  the linear system by setting $\chi_D$ and $\chi_A$ in Eqs. (\ref{nl-schrodinger}) to zero and integrating the equations of motion for a period equal to three times the system characteristic time, $T_{c}$,  defined as $1/V$.  Subsequently, we calculate the time at which the  linear system attains a  maximum in the  probability $P_A$,  $T^{max(P_A)}_{linear}$; the latter is the probability for the excitation to be at the acceptor site.  We create a PIML model using Keras \cite{Keras} as implemented in TensorFlow 2.4 \cite{TF}, with two trainable variables,  $\chi_D$ and $\chi_A$,  representing the non-linear parameters in Eqs. (\ref{nl-schrodinger}).  Initially,  $\chi_D$ and $\chi_A$,  are randomly assigned. We train the model using a custom training loop and a custom data-free physics-informed loss function defined as :

\begin{equation}
Loss(T^{max(P_A)}_{linear}, \chi_D, \chi_A) = 1 - |  \psi_{A}(T^{max(P_A)}_{linear}, \chi_D, \chi_A) |^2
\label{loss}
\end{equation}
where $| \psi_{A}(T^{max(P_A)}_{linear}, \chi_D, \chi_A) |^2$ is the probability of the system being at the targeted state, $A$,  at time,  $T^{max(P_A)}_{linear}$, for the given $\chi_D$ and $\chi_A$ values. This definition ensures that the loss function is expressed in terms of the variables $\chi_D$, and $\chi_A$ and that the the $\chi_D$ and $\chi_A$ values that minimize the loss function will maximize the occupation probability at time $T^{max(P_A)}_{linear}$. A custom training loop is then used to update the trainable parameters $\chi_D$ and $\chi_A$ of the model. In each epoch of the training loop, we integrate Eqs. (\ref{nl-schrodinger}) up to time,  $T_c$, calculate the 
maximum  probability,  $max(P_A)$, and up to time $T^{max(P_A)}_{linear}$  calculate the loss function using Eq. (\ref{loss}).  We then evaluate the gradients of Eq. (\ref{loss}) with respect to the trainable variables $\chi_D$ and $\chi_A$ and update their values using back-propagation with an adaptive learning rate.  We start with a learning rate equal to one and reduce it by 5$\%$ when the gradients change sign. A change in the sign of the gradients indicates that we have passed the minimum and thus we must now reduce the learning rate in order to avoid loosing it. Training stops when $| 1 - max(P_A)|$ is less that a threshold value, $thr_{max(P_A)}$ or when the absolute value of the mean value of the twenty last values of the trainable parameters minus the mean value of the last two values of the trainable parameters is less than a threshold value, $thr_{\chi_{D/A}}$. We used the values $thr_{max(P_A)} = 10^{-8}$ and  $thr_{\chi_{D/A}} = 10^{-4}$.  The details of the procedures followed in the method used are shown in the flowchart  (Fig. \ref{flowchart}).
\\

\subsection{Results}

\begin{figure}[ht]
\includegraphics[width=7cm]{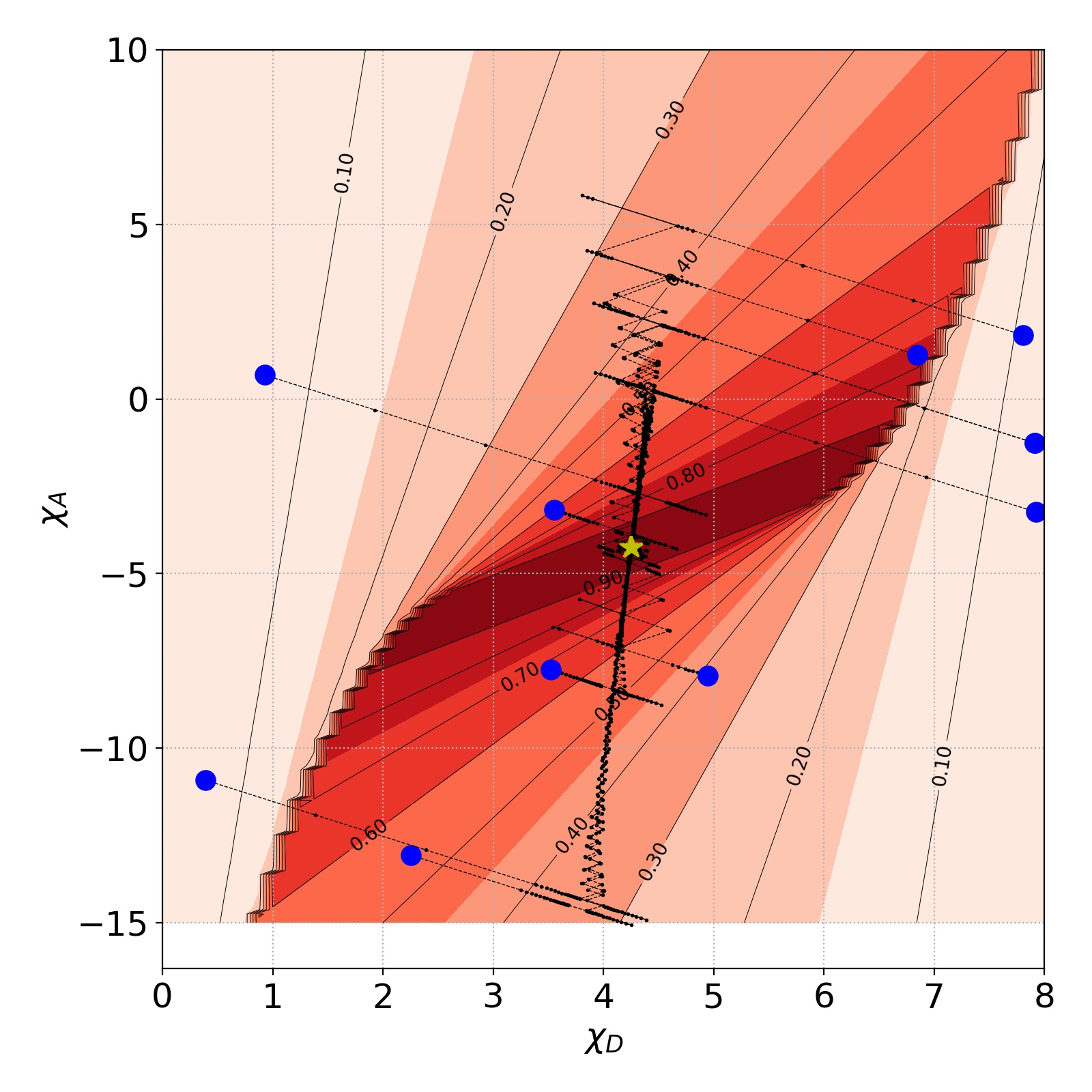} \hspace{1cm} \includegraphics[width=7cm]{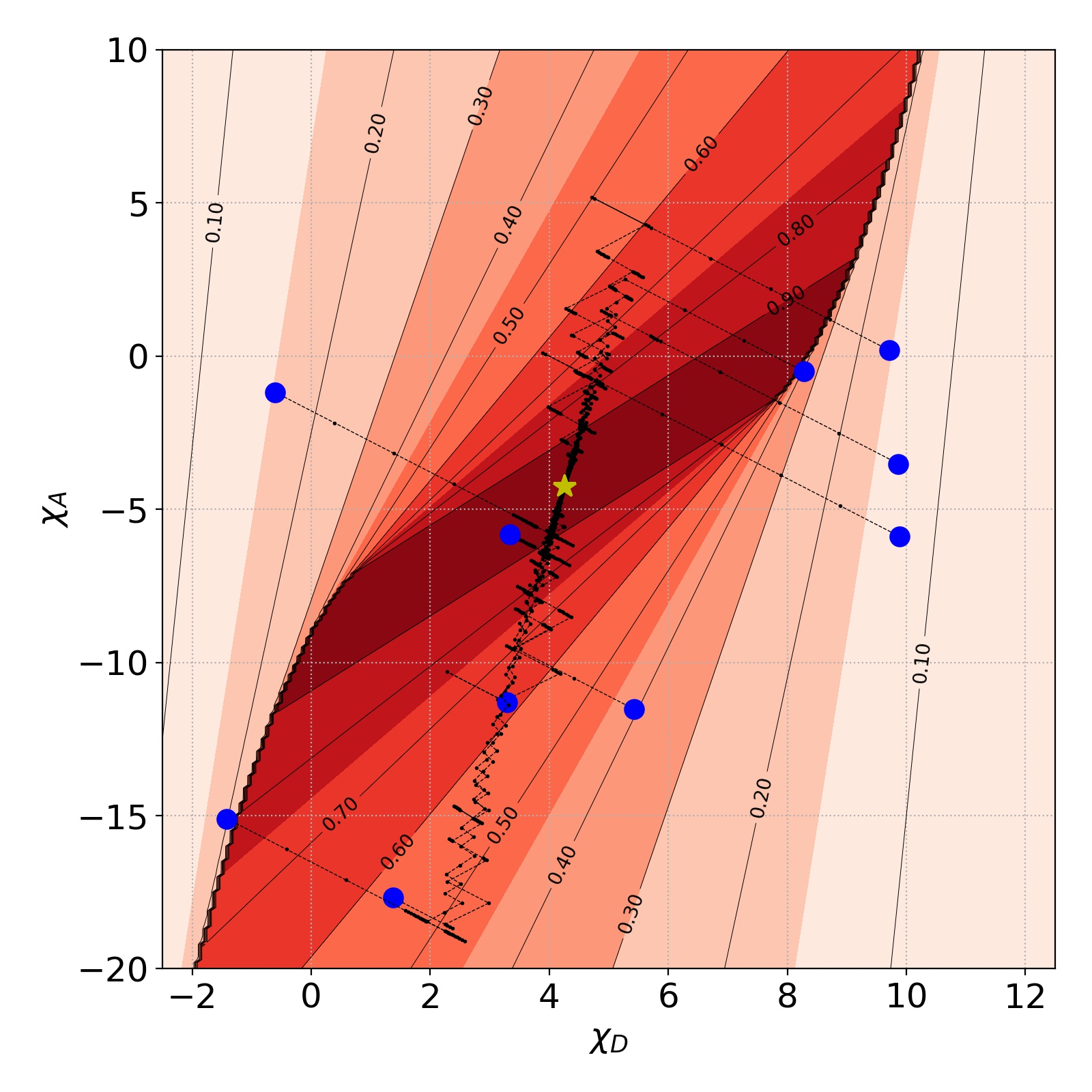} 
\caption{The occupation probability $P_A$  landscape of a non-linear Donor-Acceptor system as a function of $\chi_D$ and $\chi_A$.  Light red areas indicate low occupation probability,  dark red areas high occupation probability.  Ten randomly assigned initial values for $\chi_D$ and $\chi_A$ (blue filled circles) with their trajectories (black dashed dotted lines) and the discovered values of the parameters for each one of the trajectories (yellow star). All the trajectories conclude to the same values for $\chi_D$ and $\chi_A$.  Left: $\epsilon = 4.25$, V = 0.5. Right: $\epsilon = 4.25$ , V = 1. }
\label{DA-landscape}
\end{figure}

The back-propagation procedure we introduce is able to discover the desired dynamics based on the designed criterion and the outcome is a system with values $\chi_ D = - \chi_A \equiv \epsilon $.  The method we use was able to recover the condition that is used for targeted transfer systems (TET) \cite{kopidakis2}.  In order to explore the range and convergence to the values that give the desired transfer properties to the system, we perform a grid search where we start with different initial conditions.  In Fig. (\ref{DA-landscape}) we find the TET values for a system with $\epsilon = 4.25$ and for ten randomly assigned initial conditions (blue filled circles). In all the cases the discovering process concludes to $\chi_ D = - \chi_A = 4.25$ (yellow star).

\begin{figure}[ht]
\includegraphics[width=11cm]{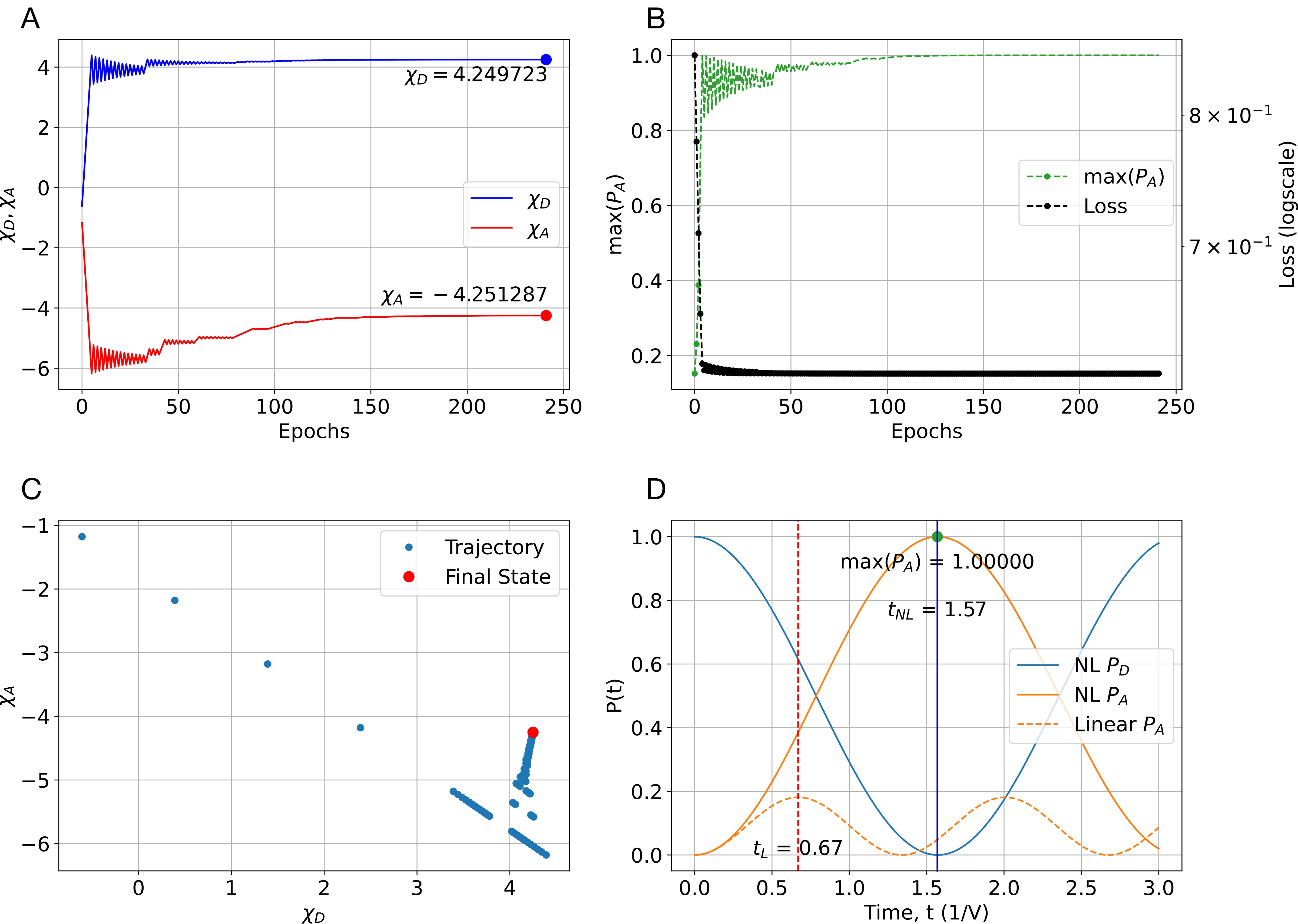}
\caption{\textbf{A. } Convergence of $\chi_D$ (blue solid line) and $\chi_A$ (red solid line) to the values with the desired transfer properties as a function of the number of epochs in the training loop.  \textbf{B. } The loss-function (black line) and the maximum occupation probability (green line) as a function of the number of epochs in the training loop.  \textbf{C. } The trajectory of the $\chi_D$ and $\chi_A$ variables during the discovering process. \textbf{D.} The occupation probability $P(t)$ at the donor (cyan), acceptor (solid orange) sites for the $\chi_D$ and $\chi_A$ values with the desired transfer properties, and the occupation probability of the acceptor (dashed orange) of the corresponding linear system as a function of time, $t$. }
\label{OneTrajectory}
\end{figure}

In Fig. (\ref{OneTrajectory}) we present the training details of one of the randomly assigned initial values of $\chi_D$ and $\chi_A$.  In the upper left graph we present the convergence of $\chi_D$ (blue solid line) and $\chi_A$ (red solid line) to the values with the desired transfer properties as a function of the number of epochs in the training loop.  In the upper right, the loss-function (black line) and the maximum occupation probability, $max(P_A)$, (green line) as a function of the number of epochs in the training loop.  The trajectory of the $\chi_D$ and $\chi_A$ variables during the discovering process are shown in the lower left graph and the occupation probability for the $\chi_D$ and $\chi_A$ values with the desired transfer properties as a function of time,  $t$ versus the occupation probability of the linear system in the lower right graph. 

\newpage

\section{Transition over a barrier into a third state}
\begin{figure}[ht]
\includegraphics[width=6cm]{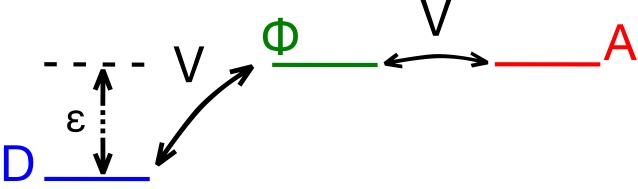}
\caption{Nonlinear trimer. Energy levels.}
\label{nl-trimer-sketch}
\end{figure}

We now turn to a more complex situation where we also use the dimer TET results found previously.  Assume that between the donor and acceptor units we have an intermediate state that for the sake of simplicity we take as completely linear as in Fig. (\ref{nl-trimer-sketch}).  A situation similar to this occurs when the donor-acceptor pair is embedded in a lattice and there is an intermediate state between them. Or, in the photonic case, a linear fiber is placed in between the two nonlinear ones. 

It is now much harder to turn the donor-acceptor system to being again resonant - the new state imposes an additional  barrier to be surmounted.   The equations of motion for the Donor-Barrier-Acceptor system are:
\begin{eqnarray}
\begin{split}
i \dot{\psi}_{D} & = & -\epsilon \psi_D + V \phi + \chi_D | \psi_D |^2 \psi_D \\
i \dot{\phi} &=& V( \psi_D + \psi_A )\\
i \dot{\psi}_{A} & = & V \phi + \chi_A| \psi_A|^2 \psi_A
\end{split}
\label{nl-schrodinger-2}
\end{eqnarray}
where we label with $\phi$ the probability amplitude for the intermediate ``barrier" state.

We apply again the discovering procedure detailed previously in analyzing the basic question, i.e. given a parametrization of the transfer rate $V$ and the energy non-degeneracy $\epsilon$, what is the donor-acceptor nonlinear values that lead to the most efficient transfer? The trimer dynamical system is in general chaotic and thus we need to search in the complex space that includes chaotic systems for the class that has the desired transfer properties.  Clearly the transfer from donor to acceptor cannot be as efficient now compared to the dimer case.  


\begin{figure}[ht]
\includegraphics[width=7cm]{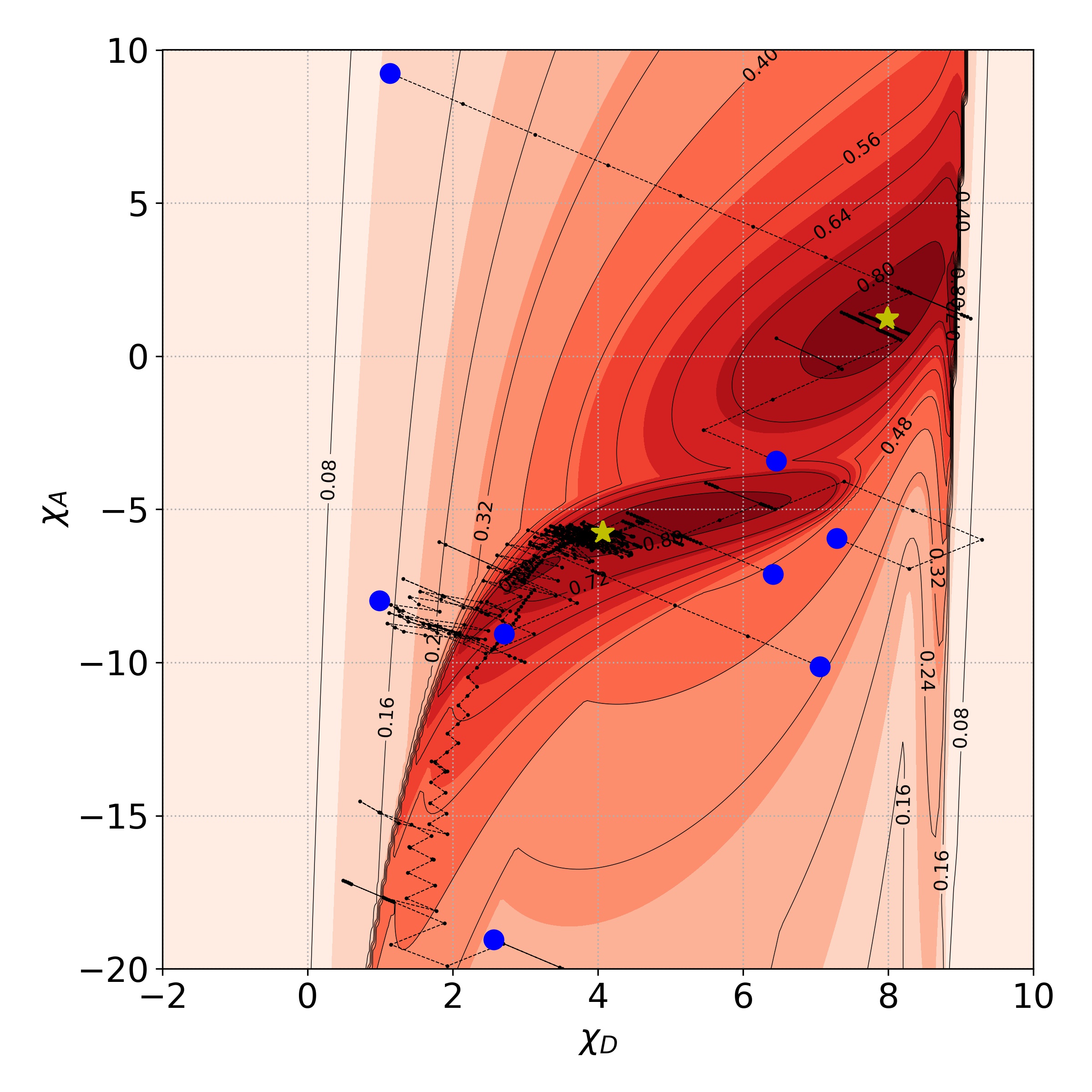} 
\caption{Donor-Barrier-Acceptor landscape with several randomly initialized (blue filled circles) trajectories (black dashed dotted lines) of the $\chi_D$ and $\chi_A$ variables during the discovering process. The star points to the position for the values of $\chi_D$ and $\chi_A$ that lead to the desired, i.e. maximum transfer.}

\label{DFA-landscape-with-trajectories}
\end{figure}

\begin{figure}[ht]
\includegraphics[width=12cm]{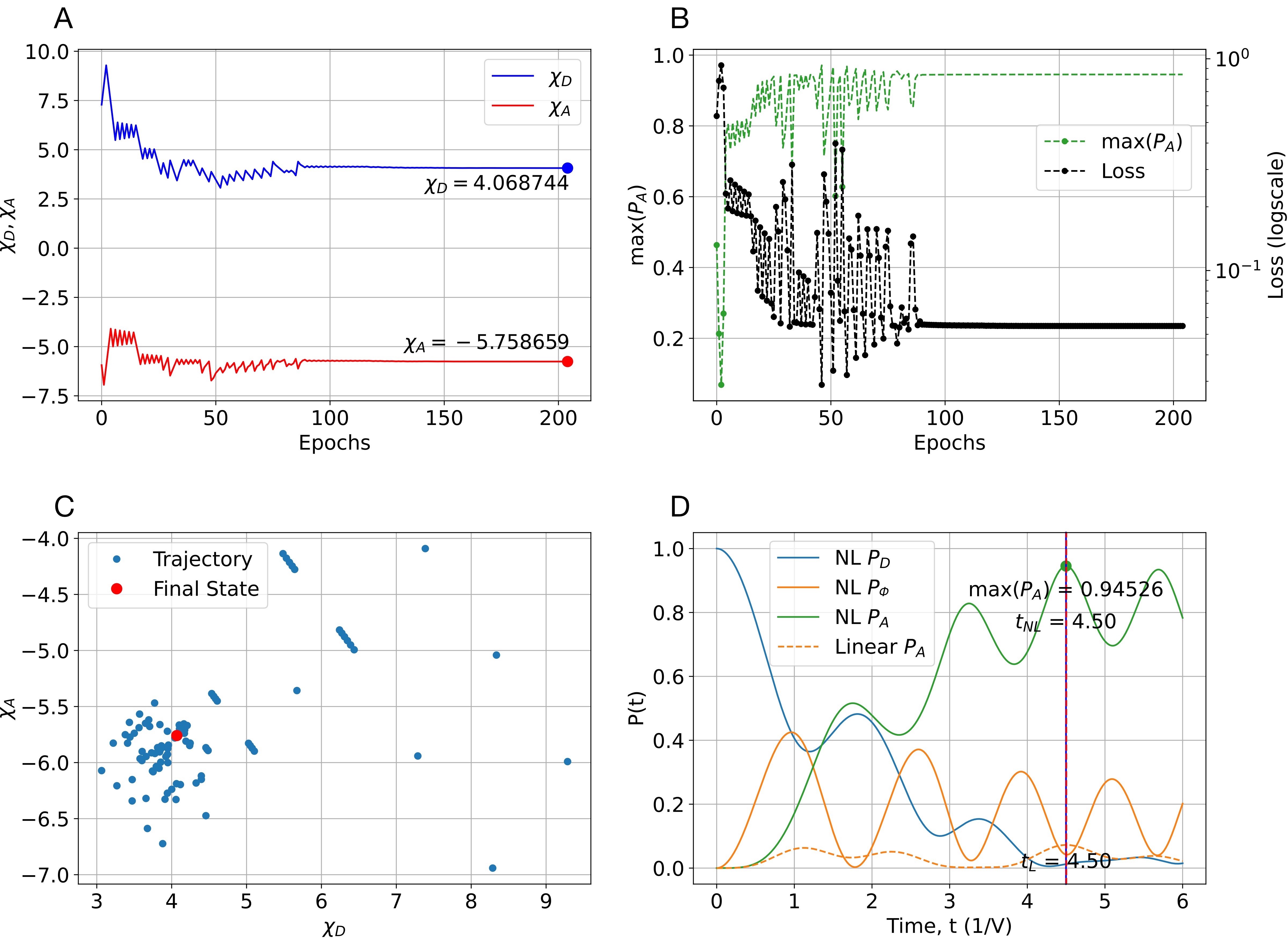}
\caption{\textbf{A} Convergence of $\chi_D$ and $\chi_A$ to the values with the desired transfer properties as a function of the number of epochs in the training loop. $\chi_D$ in solid blue and $\chi_A$ in solid red lines.  \textbf{B}.  The loss-function (black line) and the maximum occupation probability (green line) as a function of the number of epochs in the training loop.  \textbf{C} The trajectory of the $\chi_D$ and $\chi_A$ variables during the discovery process.  \textbf{D}.  The occupation probability $P(t)$ at the donor (cyan),  the intermediate (orange) and the acceptor (green) sites for the $\chi_D$ and $\chi_A$ values with the desired transfer properties, and the occupation probability of the acceptor (dashed orange) of the corresponding linear system as a function of time, $t$. }

\label{DFA_chi1chi2vsEpochs}
\end{figure}

In the case of the Donor-Acceptor system all the trajectories converge to the theoretical values of $\chi_D$ and $\chi_A$, i.e.  $\chi_D = -\chi_A = \epsilon$, where $\epsilon =  4.25$ in our case.  In the case of the Donor-Barrier-Acceptor system the occupation probability landscape is much more complex containing not only a global maximum but also local maxima and we increase the maximum integration time up to six times the characteristic time, $1/V$, of the system. Nevertheless, this method ensures that all the trajectories will conclude to the global or at least to a local maximum of the occupation probability.  In Fig. (\ref{DFA_chi1chi2vsEpochs}) the details of the discovering process of one of the trajectories that concluded to the global maximum of the occupation probability is presented. 


Choosing an adaptive learning rate helps the training process and the system converges to the parameters of the maximum occupation probability.

\newpage

\section{Discussion}

In this work we focused on the possibility to select from a class of physical systems one that has desired properties.  We worked with the standard DNLS equation and specialized it to the case of a two-state system that forms a donor-acceptor pair.  The specific form is inspired from ultra-fast electron transfer in 
chlorophyl \cite{aubry2}.  Within the class of nonlinear dimers, we searched for those with the desired electronic transfer between donor and acceptor states. We used a form of a data-free physics-informed machine learning model that through back-propagation enables efficient search in the space of systems in a way similar to the one in physics-informed neural networks.  Our back-propagation procedure converges fast and recaptures the TET condition found analytically  earlier \cite{kopidakis2}.  This is a very significant result since it shows that proper use of physics-motivated machine learning may lead to a ``discovery" that has solid mathematical basis.  In other words,  it appears that ML has the capacity to find established results in dynamical systems.

Having searched and actually found the configuration that concludes to the desired transfer properties for the simple donor-acceptor system we can go beyond it and try to discover configurations that are either unknown or difficult to asses.  We thus extended the method to the case where a barrier in the form of an additional linear site was inserted in between the donor-acceptor pair.  It is not easy to find analytically, or even numerically through nonlinear resonance search the systems with designed transfer \cite{kopidakis2}.  However, our method works well and is able to discover efficiently the systems with the desired properties.  The method, in addition to molecular and optical systems may be applied also to mechanical ones where the concept of TET and the engineering use of nonlinear resonances proves to be particularly interesting \cite{vakakis}.

Our findings, although applied in a small set of systems, are of more general scope.  We believe that the method we introduce here can be used in order to construct nonlinear dynamical systems with prescribed properties without having to dwell  on the complex resonance structure of these systems.  Furthermore, since the DNLS is a semi-classical equation that describes quantum mechanical transfer one may also apply our method to strongly interacting quantum systems and be able to design photonic as well as quantum meta-materials with specific properties \cite{soukoulis, zagoskin}.

\section{Disclosures}
The authors declare no conflicts of interest.

\section{Data availability.}
No data were generated or analyzed in the presented research.

\begin{acknowledgments}
We thank Serge Aubry for helpful discussions. This work was partially supported  by the Institute of Theoretical and Computational Physics of the University of Crete and also by the Ministry of Science and Higher Education of the Russian Federation in the framework of Increase Competitiveness Program of NUST "MISiS" (No. K2-2019-010), implemented by a governmental decree dated 16th of March 2013, N 211. 
\end{acknowledgments}

\end{document}